# Analysis of some solid amorphous inorganic structures and the boson peak phenomenon with a computational random graph approach

A. Berezner [a*], M. Rybakov [b], M. Sidlyar [c], V. Fedorov [a]

[a] Theoretical and Experimental Physics Department, Derzhavin Tambov State University, Tambov 392000, Russia

[b] Functional Analysis Department, Derzhavin Tambov State University, Tambov 392000, Russia

[c] Mathematical Modeling and IT Department, Derzhavin Tambov State University, Tambov 392000, Russia

[*]E-mail for correspondence: a.berezner1009@gmail.com

**Abstract**

In this study, a new alternative model algorithm has been proposed for assembling amorphous structures, unifying the bosonic paradigm applicable at low temperatures with crystalline models relevant at room and higher temperatures. Physical meaning of main model parameters is determined together with an explanation for the appearing bosonic peak using the random graph theory. Numerically, statistical atomic distribution in a multiphase amorphous system is provided without the melting simulation of base crystals, and the mean energy function has been determined analytically. The calculated table data are in good agreement with neutronography measurements of the actual amorphous alloy in its solid state. Programme optimisations were also implemented, and we outlined several effective steps to achieve the higher processing speed. The proposed programme code can be used for potential test assembling and simulations of amorphous systems with sorting by the optimal atomic content or proportion (i.e. glass forming ability).

# 1. Introduction

Amorphous metallic alloys (also known as metallic glasses, MGs) and other disordered systems are under intensive study since their first synthesis [1,2]. Besides their unique applied properties caused by a non-crystalline structure [3,4], understanding of key principles driving an amorphous molecular assembling is interesting [5,6], and there are also unexplored effects [7,8] for these materials.

A notable feature typical for metallic glasses and other disordered systems is the bosonic peak [9], which is absent in crystalline materials and appears in thermal capacity plots at cryogenic temperatures [10-12]. Despite the qualitative understanding of the phonon (Bose-Einstein) statistics in MGs, there is no main model relationship between disordered real structures and their thermodynamic parameters in the whole temperature interval (from cryogenic to crystallisation point) yet. As a source of the bosonic peak in metals, pair atomic dipoles (also dumbbells) [13] or crystal-like interstitial defects [14], possibly frozen in the amorphous matrix during its casting at $10^3$-$10^6$ K/s cooling rates [15], are supposed. However, their physical presence in metallic glasses and spatial distribution relative to ordered structures (at crystallisation of amorphous alloys, for instance) remain unclear. Also one should note the transition from a bosonic statistic of the whole amorphous system into the normal or more complex distribution at the room and higher temperatures [17,18], and that can be possible for the phonons [19] and molecules. Furthermore, in the review [16], it is noted that for a complete understanding of structural formation in amorphous alloys, novel computational approaches are necessary.

Other attractive properties of metallic glasses (such as high longitudinal strength and low magnetisation work) are also modelled with respect to structural properties [20,21] depending on a crystalline basement. For example, at room temperatures or above, combined approaches like crystal melting [22], embedded atom (EAM) [23], density functional (DFT) [24], phenomenological or semi-empirical potentials [25], and molecular dynamics [26] are commonly used. At the first computational step, *N*-particle single crystals are modelled alongside the sum of their fundamental energy cells (such as Voronoi and Casper polyhedra [27] or Bravais lattices [28]). Then, using the first thermodynamic law [29] and total energy conservation principle [30], the temperature value is changed up to reaching the empirical melting point with proportional reverse compensations by «atomic» coordinates (material points) for maintaining constant total energy in the system (i.e. one changes several variables in the isoenergetic functional layer surface). After that, coordinate arrays are randomly rewritten in the loop at the constant energy to the complete temperature return (the cooling part) without a full match in the spatial numbers (here we mean Andersen, Berendsen, and other algorithms, which are extensions of the Nose-Hoover thermostat [31]). In result, melting and cooling numerical procedures provide assembling of non-crystalline arrays as the order is eliminated with isoenergetic cycles, and accompanying thermodynamic parameters like Young's modulus, viscosity, distribution (PDF), etc. can be calculated for amorphous structures. Herewith, two- or three-particle base crystalline lattices are usually considered [32], and their energies of a primitive unit cell are main parameters in density functional theory [33], eigenvalues in the Schrodinger equation [34], or even collective quantities in many-body approximations [35]. As a disadvantage of the mentioned approach, the mandatory relation between an amorphous structure and its crystal pseudo-domain (i.e. prehistory in the unit cell) must be accounted. Multicomponent analysis (3 and more fractions) becomes complicated due to the presence of several ordered isolated phases as real metallic glasses are made with combined melting of separated crystalline alloys. Moreover, choosing an appropriate master potential for multicomponent metallic glasses is a complex challenge, since it must account for a system, containing numerous isolating boundaries and crystalline interphases, breaking during the heating and cooling (see Maxwell's «demon» thought experiment [36]).

An alternative necassary approach involves developing a new algorithm that unifies low-temperature bosonic statistics (dominated by atomic pairs or phonons) with collective particle interactions at room temperature [37-39], and this method should be validated through experimental investigations with real heating, viscoelastic deformation, or fracture processes [18]. Herewith, optimal description of amorphous structures with a free component number, independent from prior crystalline structures, is quite important. Also for computational optimisation, the novel programme approach should provide both bond enumeration on a single processor [40] and simplified parallel computing utilising isolated CPUs or GPUs [41].

Thus, the goal of this study is to develop a novel programming algorithm to combine different model approaches for amorphous inorganic structures by their atomic composition with respect to pair potentials, computational peculiarities, and the system tendency to the energy minimum. Besides the computational part, comparison between numerical and experimental data is realised here together with literature analysis. Additionally, the computational complexity of the proposed algorithm is evaluated along with potential optimisation strategies (such as parallel processing, list nesting, and code porting) to enhance efficiency.

## 2. Materials and methods

To achieve the stated goal, a novel Python user programme code has been developed (see Appendix, A.2). As a model computational basement, we used Finnis-Sinclair pair potentials [42] to simulate collective particle interactions in the metallic glass. The AMAG-225 amorphous alloy (at.wt. Fe-73.5%, Ni-25%, Cr-1.5%) was chosen as a benchmark system with pair metal potentials adopted from existing literature (for instance, [43] and others). Analytical expressions for pair potentials are provided in section (A.1) of the Appendix. The calculations were performed for $10^4$ particles mainly in 2D space using JSCC RAS hardware (see Acknowledgements) specifically two nodes equipped with 96 Intel Xeon Platinum 8268 processors each. Furthermore, we made calculations either with four isolated processors [40], or using parallel coding on the whole allowed structure by random point distribution (as mentioned in the section ii) of (A.2)). Computed data were compared with neutron scattering (the DN-2 thermal neutron diffractometer, JINR) on amorphous alloys in the solid state by the method described in [44] or its references, and we interpolated plot points statistically using the linear correlation coefficient (for instance, 0.99 at 20 energy tables) with the radial distribution function. Separately, we analysed disordered atomic packing within the framework of the random graph theory [45], along with its relevant physical interpretation.

**3. Results and discussion**

Let us consider the programme workflow, detailed in the Appendix section, focusing on the physical meaning and technical features. In the specified spatial range (a «box» with potential particles), point sets of coordinate pairs (triplets in the volumetric case) were randomly placed without repeating at every new programme launch, and their total amount was proportionally equal to percent atomic composition in the AMAG-225 alloy. Then, interatomic distances between all the possible points (as potential atomic centres) were calculated in loops up to the last pair indexes. Since permutation distances (with *ij* or *ji* indices) repeated within the sorting loops, duplicates were removed. Next, all the interatomic distances, obtained as a numeric array, are substituted into the pair potentials for all possible bonds to determine the minimal value, which corresponds to the binding energy. Herewith, the Finnis-Sinclair potential was used for model metallic bonding due to a hypothesis on residual pair atomic interaction in the metallic glass after the melt solidification. Unlike Einstein's model, using for description of amorphous alloys at cryogenic temperatures [9], our pair binding energy is determined not by a constant ($h\omega/2\pi$) phonon energy, but with Finnis-Sinclair values, depending on particle types in every atomic pair (see A.1). Our code also accounts for the possible presence of duplicate distances and energies between other different atoms and the same one (i.e. two or more duplicates). In this case, code executing terminates with the user error message indicating about detected duplicates. It indicates the ordering (i.e. possible local crystallisation) in a random amorphous matrix. Such approach is useful for estimation of a glass-forming ability at given atomic compositions in alloys, i.e. the system will tend to crystallise if duplicates appear. Sure, in the code, one-time use of the particle pair is considered, i.e. atomic amount decreases by one or two depending on formed dipole types after the *i*-th loop energy minimum sorting. In result, arrays, containing interatomic distances and corresponding minimum binding energies, are stored either in a text file or as an image (for example, see Fig. 1).

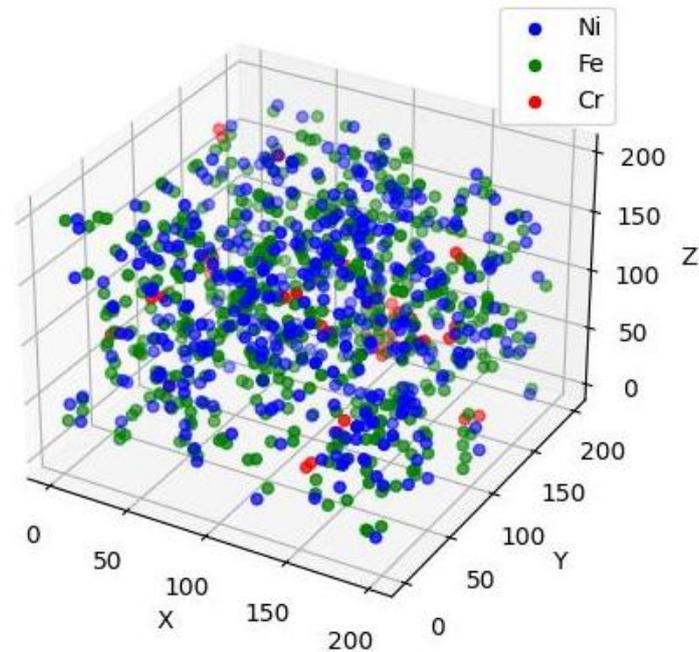

Fig.1. A typical graphical output for a spatial atomic assembling of 2000 particles in the modelled Fe-Ni-Cr amorphous structure

Thus, the programme is based on the preliminary random point distribution with the further choice (filtering) of the pairs, providing the energy minimum in the generated distances. Moreover, during the array sorting of coordinate numbers with corresponding energies, the found minimum determines both a particle type (the point gets atomic properties) and the most possible bonding pair. In other words, random point sets are generated and considered with the further selection of appropriate ones, corresponding to the most stable energy condition depending on atomic types and their proportions.

Algorithm complexity analysis testifies to the polynomial spend time with a higher amount of particles (see Fig.2) as we used nested «*for*» loops and element searching in the lists (i.e. the «*if x* in *list*» construction) [46].

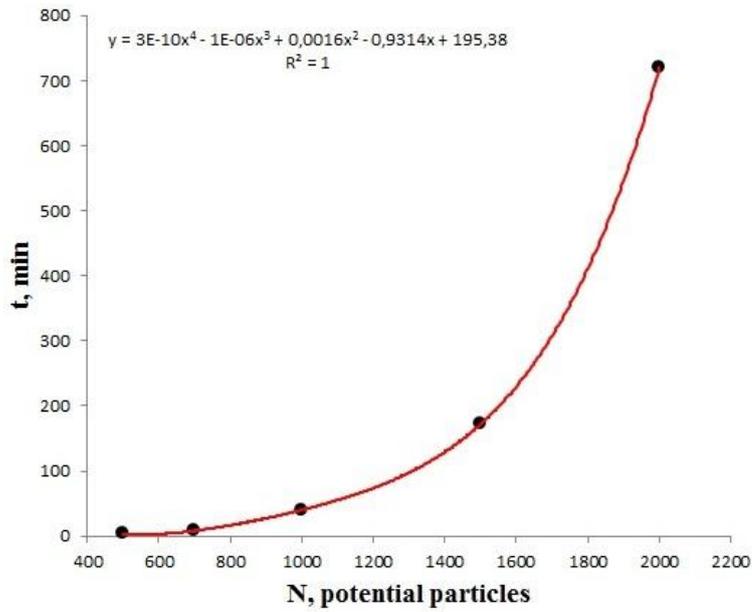

Fig.2. Empirical relationship between the amount of particles and computational time

As possible optimisation ways, enable parallel computing module (MPI) or using more Python compact arrays (the *array* class), hardware GPU or CPU parallel working, and porting to other languages (like *C*, for instance) could be used. Therefore, we optimised our code using MPI for dynamical processor resources as deemed the system growth by scale (see ii) of (A.2) in Appendix), and also compared it with the same calculations but on isolated fixed CPU kernels respected to Plimpton's recommendations [40]. These steps provided 19 times or more high calculation speed with an expansion of the upper particle limit from 2000 to $10^4$. Herewith, both approaches (MPI and isolated CPU computing) led to the same result at plotting of the U(r) radial distribution function for the model amorphous system (like in Fig.3). As shown in Fig. 3a, stable cyclic operation is achieved with high accuracy, indicated by a 0.99 linear correlation coefficient and ±0.011 eV absolute error (Fig. 3d). And the output of separated phase energies (Fig. 3b) is possible together with the identity between the total energy curve and neutron scattering data on solid MGs by 5.6-5.8 angstrom (A) interatomic distance with asymptotic drop (at 10 A), and matching the whole shape (Fig. 3c).

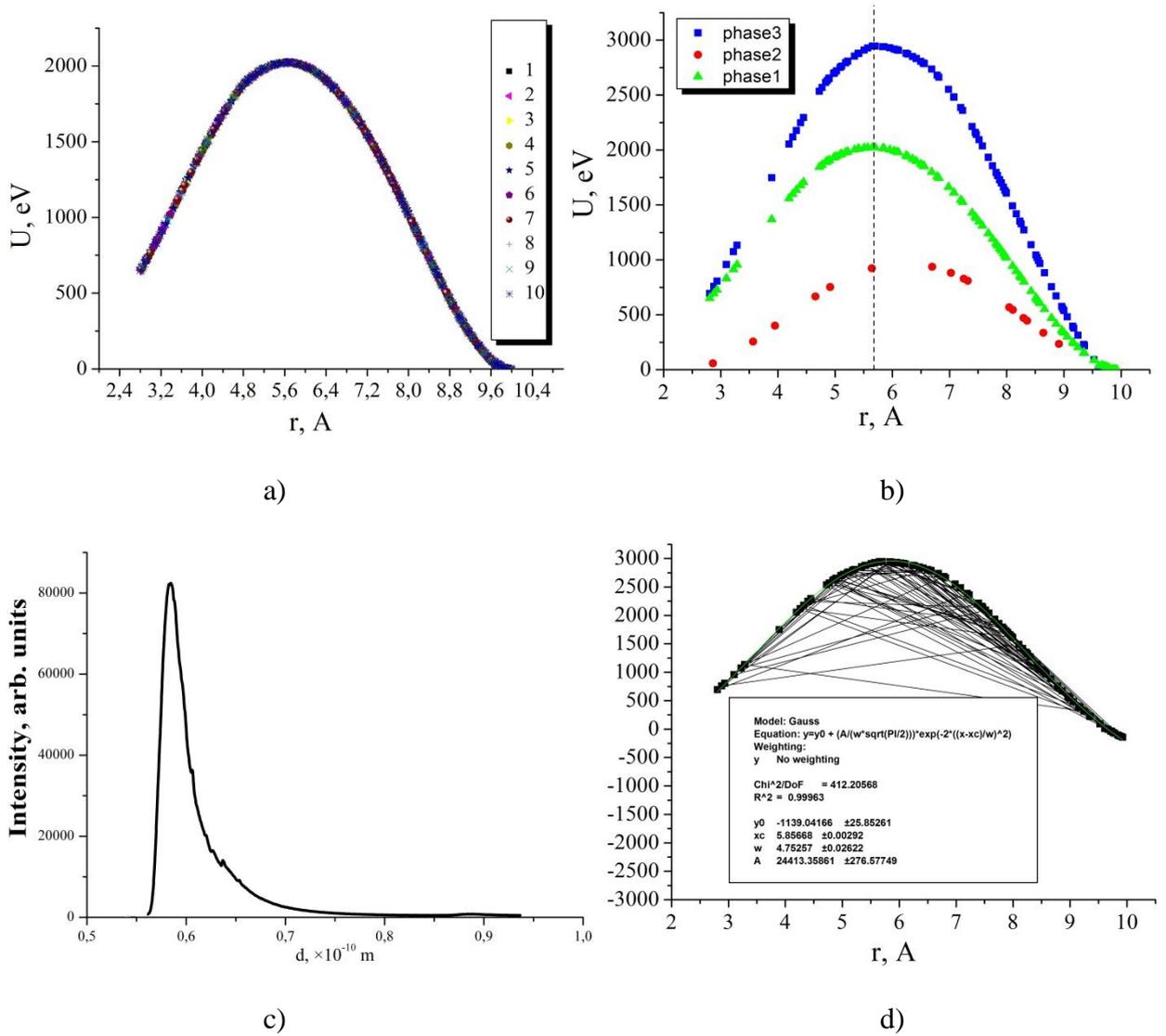

Fig.3. Calculated U(r) function for the AMAG-225 MG: a) plotted by 10 programme launches (i.e. every marked point type corresponds to a full-fledged separate launch); b) numerical metal phase separation during a launch; c) experimental neutron scattering data (to compare); d) statistical distribution of the numerical atomic bonding (graph points are linked with black lines to demonstrate a random behaviour during pair assembling and curve plotting)

Data interpolation testifies to the Gauss energy distribution for disordered atomic bonding regardless of the plotting point sequence (Fig. 3d).

Considering that mentioned calculations can be interpreted as pair vertex matching in a random graph (always following a Poisson or Gauss distribution [45]), we can prove the robustness of the used pair Finnis-Sinclair potentials here. For this, it is enough to accept $<U(x)>$ as a Gauss exponent in the mean energy function expression $<U(x)> = \int U(x)P'(x)dx$ (1), considering the curve distribution (Fig. 3b, d), together with a substitution of a Finnis-Sinclair function from i) (see Appendix) instead of $U(x)$ in the integral and further calculation of $P'(x)$. Then if we differentiate (1) with the further expression of $P'(x)$, a Gauss distribution and the polynomial Finnis-Sinclair multiplier like ones in [45], dedicated to a random graph analysis on any vertexes, edges, and matching, will be derived. In [45], the expression for the increasing matching number on random edges, which corresponds to the relaxation $β$-peak function in metallic glasses [9] (i.e. a KWW stretched exponent), is also given. It permits conclusions about possible modelling of amorphous structures using random graph methods with matching numbers and their changing probability at forming the ordered (crystallisation) or stochastic (thermal relaxation) complex chains. Calculated arrays correspond to block adjacency matrices with an acting operator. Since the original Finnis-Sinclair potential is considered as an atomic binding energy in crystalline cells [42], its application for MGs becomes meaningful in the following context here. Preserved after the melting of a nominal crystal, pair interatomic bonds (i.e. randomly distributed edges) form a solid amorphous structure (the random graph with pair-matched vertexes), whose energy is proportional to $exp(-R^2)$ or $exp(-E/kT)$ functions with respect to the canonical Gibbs ensemble (where $k$ is the gas constant, $T$ is absolute temperature, and $E=U(x)$ represents pair atomic potential, and $R$ is a distance or local (also mean) free path). Then differentiation of (1) by $T$ (at $-E/kT$ representation) provides the expression for system thermal capacity $c \sim exp(-E/kT)/T^2$, having the local minimum in the low temperature region (i.e. the bosonic peak [13,16]). The magnitude of a bosonic peak depends on constant parameters (bond rigidity and atomic types) in Finnis-Sinclair potentials at zero external perturbations (i.e. without phonon modes or excess thermal photonic fluxes), and only frozen atomic pairs provide such bosonic response during cooling in the cryogenic region. In the room or higher temperatures, discrete Poisson's or smooth Gauss behaviour predominates [45], i.e. initially isolated pair cryogenic contribution becomes into the broad action (a completely connected graph) like a biatomic collective ideal gas, which is modelled with Johnson-Mehl-Avrami-Kolmogorov (JMAK) [47], Kohlrausch-Williams-Watts (KWW) [9], and other thermal models [18]. Since graph matching occurs not only between pairs [45] without altering the Gaussian statistics, the collective interaction of particles (such as relaxation and crystallisation) can be viewed as a transition from a double-vertex structure to an N-point directed network in an amorphous material during heating.

## 4. Conclusion

Considered disorder structural assembling universally simulates formation of the metallic glass in adiabatic and stationary approximations. Time accounting for the elementary pair particle interaction could be provided using the Python system clock module, but this calculation depends on semiconductor processes in a CPU chip, and the measure of $t$ time difference with $E$ energy is ineffective considering Heisenberg's indeterminacy principle ($\Delta t \Delta E \sim h/2\pi$). In the proposed computational approach, various pair potentials can be used without model melting of the original crystalline lattices, phases, or components, but comprehensive data on atomic bonding in rare-earth elements remains still incomplete. To further enhance the code performance, porting to *C*-languages may be beneficial, as their arrays are stored separately from metadata and do not initialise with each element of a list every time. The obtained result for the mean system energy can be combined with non-isothermal data and models from [18,48], enabling a unified description for the behaviour of metallic glasses across low, room, and higher temperature regimes.

**Funding:** This research did not receive any specific grant from funding agencies in the public, commercial, or not-for-profit sectors.

**Declaration of competing interest**

The authors declare that they have no known competing financial interests or personal relationships that could have appeared to influence the work reported in this paper.

**Acknowledgements**

**Appendix**

(A.1) Atomic pair bonding in AMAG-225

```
The Cr-Cr pair:

If 0<x<10 angstrom, then U(x)=((x-10)^2)*(29,1429813-
(23,397503*x)+(4,7578297*(x^2)));
If x>10 angstrom, then U(x)=0;
-------------------------------------------------------------------------------
The Fe-Fe pair:

If 0<x<10 angstrom, then U(x)=((x-10)^2)*(26,27034-
(24,40109*x)+(6,957871*(x^2)));
If x>10 angstrom, then U(x)=0;
-------------------------------------------------------------------------------
The Ni-Ni pair:

If 0<x<10 angstrom, then U(x)=((x-10)^2)*(13,28276-
(17,08506*x)+(8,262515*(x^2)));
If x>10 angstrom, then U(x)=0;
-------------------------------------------------------------------------------
The Fe-Cr pair:

If 0<x<10 angstrom, then U(x)=((0,850704*(x-10)^2)*(26,27034-
(24,40109*x)+(6,957871*(x^2))))+((0,186420*(x-10)^2)*(29,1429813-
(29,3975027*x)+(4,7578297*(x^2))));
If x>10 angstrom, then U(x)=0;
-------------------------------------------------------------------------------
The Fe-Ni pair:

If 0<x<10 angstrom, then U(x)=((0,516721*(x-10)^2)*(26,27034-
(24,40109*x)+(6,957871*(x^2))))+((0,275781*(x-10)^2)*(13,28276-
(17,08506*x)+(8,262515*(x^2))));
If x>10 angstrom, then U(x)=0;
-------------------------------------------------------------------------------
The Ni-Cr pair:

If 0<x<10 angstrom, then U(x)=((0,80590965*(x-10)^2)*(13,28276-
(17,08506*x)+(8,262515*(x^2))))+((0,192553163985*(x-10)^2)*(29,1429813-
(23,3975027*x)+(4,7578297*(x^2))));
If x>10 angstrom, then U(x)=0;
```

## (A.2) The main programme code

### i) *Model assembling of the AMAG-225 alloy*

```python
print('Disordered atomic assembling')
import math
import random
import copy
import datetime
import matplotlib.pyplot as plt
a1=4000     #a size of the «box» with particles (minimal)
b1=6000     #a size of the «box» with particles (maximal)
N1=1460     #-|
N2=500      #-| Amount of particles is set by types proportionally (Fe-73%,
Ni-25%, Cr-1.5%) considering computational abilities
N3=40
N=N1+N2+N3
S1=[]
S2=[]
for i in range(0,N): #Amount of particles is determined with the range()
function, and the upper limit must be N1+N2+N3
    a=round(random.uniform(a1,b1),3) # the «3» number inside the uniform()
function corresponds to coordinate precision
    b=round(random.uniform(a1,b1),3)
#-|===> For the 3D count, a c=round(,,) expression should be set instead this
comment
    S1.append(a)
    S1.append(b)
for i in range(0,len(S1),2): #For 3D count, the «3» number must be input here
and in the next line instead the «2»
    S2.append(S1[i:i+2])
print('Possible coordinates') #Here is created the list of random coordinate
pairs (or triplets) for potential atomic point centres (Xo;Yo) or (Xo;Yo;Zo)
T1=datetime.datetime.now()
print(T1,'\n')
L=[]
for i in range(0,len(S2)): # Cyclic calculation of primary interatomic
distances between possible atoms (with full coordinate sorting).
    for n in range(0,len(S2)):
        L.append(round(math.sqrt((((S2[n][0]-S2[i][0])** 2)+((S2[n][1]-S2[i][1])** 2))),3))
S4=[]
for i in range(0,len(L),len(S2)):
    S4.append(L[i:i+len(S2)])
print('Primary interatomic distances')
T2=datetime.datetime.now()
print(T2,'\n')
S3=[]
S5=[]
for i in range(0,len(S4)):
    for k in range(0,len(S4[i])):
        if S4[i][k] in S5:
            S4[i][k]=0
        else:
            S5.append(S4[i][k])
print('Interatomic distances without duplicates')
T3=datetime.datetime.now()
print(T3,'\n')
for i in range(0,len(S4)):
```

```python
        for k in range(0,len(S4[i])):
            if S4[i][k]<2.7: #Filtration of distances less than double minimal atomic radius (2*R(Ni)~ 2*1.35 angstrom), and zeros are also neglected.
                S4[i][k]=0
            else:
                None
print('Filtered interatomic distances')
T4=datetime.datetime.now()
print(T4,'\n')
B1=copy.deepcopy(S4) #Making a sorted list with minimal energies (as an operator image of distances)
for i in range(0,len(B1)):
    for j in range(0,len(B1[i])):
        B1[i][j]=0
M1=[] #The empty and renewable sorting buffer for the different energy numbers
C1=N1
C2=N2
C3=N3
for i in range(0,len(S4)):
    for k in range(0,len(S4[i])):
        if 2.7<=S4[i][k]<=10: #Check of sorted distances for the maximal neutron diffraction limit (10 angstrom cross section)
            if 2.8<=S4[i][k]<=10 and C3>=2: #Check if there are particles for Cr-Cr bonding
                M1.append(round(((S4[i][k]-10)**2)*(29.1429813-(23.397503*S4[i][k])+(4.7578297*(S4[i][k]**2))),3)) #Binding energy for the Cr-Cr pair
            else:
                M1.append(0)
            if 2.8<=S4[i][k]<=10 and C1>=2: #Check if there are particles for Fe-Fe bonding
                M1.append(round(((S4[i][k]-10)**2)*(26.27034-(24.40109*S4[i][k])+(6.957871*(S4[i][k]**2))),3)) #Binding energy for the Fe-Fe pair
            else:
                M1.append(0)
            if 2.7<=S4[i][k]<=10 and C2>=2: #Check if there are particles for Ni-Ni bonding
                M1.append(round(((S4[i][k]-10)**2)*(13.28276-(17.08506*S4[i][k])+(8.262515*(S4[i][k]**2))),3)) #Binding energy for the Ni-Ni pair
            else:
                M1.append(0)
            if 2.8<=S4[i][k]<=10 and C1>=1 and C3>=1: #Check if there are particles for Fe-Cr bonding
                M1.append(round(((0.850704*(S4[i][k]-10)**2)*(26.27034-(24.40109*S4[i][k])+(6.957871*(S4[i][k]**2))))+((0.186420*(S4[i][k]-10)**2)*(29.1429813-(29.3975027*S4[i][k])+(4.7578297*(S4[i][k]**2)))),3)) #Binding energy for the Fe-Cr pair
            else:
                M1.append(0)
            if 2.75<=S4[i][k]<=10 and C1>=1 and C2>=1: #Check if there are particles for Fe-Ni bonding
                M1.append(round(((0.516721*(S4[i][k]-10)**2)*(26.27034-(24.40109*S4[i][k])+(6.957871*(S4[i][k]**2))))+((0.275781*(S4[i][k]-10)**2)*(13.28276-(17.08506*S4[i][k])+(8.262515*(S4[i][k]**2)))),3)) #Binding energy for the Fe-Ni pair
            else:
                M1.append(0)
            if 2.75<=S4[i][k]<=10 and C2>=1 and C3>=1: #Check if there are particles for Ni-Cr bonding
```

```python
                        M1.append(round(((0.80590965*(S4[i][k]-10)**2)*(13.28276-
(17.08506*S4[i][k])+(8.262515*(S4[i][k]**2))))+((0.192553163985*(S4[i][k]-
10)**2)*(29.1429813-(23.3975027*S4[i][k])+(4.7578297*(S4[i][k]**2)))),3))
#Binding energy for the Ni-Cr pair
                    else:
                        M1.append(0)
                M2=[]
                for j in range(0,len(M1)): #Deleting of zeros (if they are)
                    if M1[j]!=0:
                        M2.append(M1[j])
                    else:
                        None
                M1=M2
                try:
                    if len(set(M1))==len(M1):
                        None
                    else:
                        print('Detected duplicates! Please, restart the
programme')
                        exit()
                except ValueError:
                    None
                try:
                    B1[i][k]=min(M1) #List filling/rewriting by minimal energy
numbers
                    if min(M1)==round(((S4[i][k]-10)**2)*(29.1429813-
(23.397503*S4[i][k])+(4.7578297*(S4[i][k]**2))),3):
                        C3=C3-2
                    elif min(M1)==round(((S4[i][k]-10)**2)*(26.27034-
(24.40109*S4[i][k])+(6.957871*(S4[i][k]**2))),3):
                        C1=C1-2
                    elif min(M1)==round(((S4[i][k]-10)**2)*(13.28276-
(17.08506*S4[i][k])+(8.262515*(S4[i][k]**2))),3):
                        C2=C2-2
                    elif min(M1)==round(((0.850704*(S4[i][k]-10)**2)*(26.27034-
(24.40109*S4[i][k])+(6.957871*(S4[i][k]**2))))+((0.186420*(S4[i][k]-
10)**2)*(29.1429813-(29.3975027*S4[i][k])+(4.7578297*(S4[i][k]**2)))),3):
                        C1=C1-1
                        C3=C3-1
                    elif min(M1)==round(((0.516721*(S4[i][k]-10)**2)*(26.27034-
(24.40109*S4[i][k])+(6.957871*(S4[i][k]**2))))+((0.275781*(S4[i][k]-
10)**2)*(13.28276-(17.08506*S4[i][k])+(8.262515*(S4[i][k]**2)))),3):
                        C1=C1-1
                        C2=C2-1
                    elif min(M1)==round(((0.80590965*(S4[i][k]-10)**2)*(13.28276-
(17.08506*S4[i][k])+(8.262515*(S4[i][k]**2))))+((0.192553163985*(S4[i][k]-
10)**2)*(29.1429813-(23.3975027*S4[i][k])+(4.7578297*(S4[i][k]**2)))),3):
                        C2=C2-1
                        C3=C3-1
                    else:
                        None
                except ValueError:
                    None
                M1=[]
                if B1[i][k]!=0:
                    break
            else:
                B1[i][k]=0
print('Preliminary energy numbers') #Rewriting of distances by energies on
the same positions
T5=datetime.datetime.now()
print(T5,'\n')
U=[] #Making the filtered list of one-time use pair bonding energies.
```

```python
for i in range(0,len(B1)):
    B2=[]
    for k in range(0,len(B1[i])):
        if B1[i][k]!=0:
            B2.append(B1[i][k])
        else:
            None
    if len(set(B2))==len(B2):
        None
    else:
        print('Detected duplicates! Please, restart the programme')
        exit()
    try:
        U.append(min(B2))
    except ValueError:
        None
    for l in range(0,len(B1)):
        for m in range(0,len(B1[l])):
            if B1[l][m] in U:
                k=B1[l].index(B1[l][m])
                B1[k]=[0] # Deleting of excess energies
                for t in range(0,len(B1)):
                    for c in range(0,len(B1[t])):
                        if c==k:
                            B1[t][c]=0
            else:
                None
print('Final (useful) pair bonding energies')
T6=datetime.datetime.now()
print(T6,'\n')
#============================ Graphic block

#Particle plotting by coordinates in a user function
def MyPlot(S7, K):

    #Generate the figure with markers
    fig = plt.figure('Point generation')
    ax = fig.add_subplot(111)
    #Mark the X axis
    ax.set_xlabel('X')
    #Mark the Y axis
    ax.set_ylabel('Y')

    #open the graphical window
    plt.get_current_fig_manager().window.state('zoomed')

    #The particle list
    atoms = []
    for a in S7:
        patoms = a.split('-')
        for b in patoms:
            atoms.append(b)

    #The list of particles
    typeAtoms = list(set(atoms))

    #Particle amount
    nAtoms = len(set(atoms))

    #Particle counter
    i = 0

    #Particle coordinates in list
```

```python
    coordAtoms = [[[],[]] for _ in range(nAtoms)]

    #Get coordinates for different particle types
    while i < nAtoms:
        j = 0
        for x,y in K:
            if typeAtoms[i] == atoms[j]:
                coordAtoms[i][0].append(x)
                coordAtoms[i][1].append(y)
            j += 1
        i += 1

    #Colored particles coordinates by types
    n = 0
    while n < nAtoms:
        if typeAtoms[n] == 'Cr':
            ax.scatter(coordAtoms[n][0],coordAtoms[n][1],
                       label=typeAtoms[n], c="cyan")
        if typeAtoms[n] == 'Ni':
            ax.scatter(coordAtoms[n][0],coordAtoms[n][1],
                       label=typeAtoms[n], c="black")
        if typeAtoms[n] == 'Fe':
            ax.scatter(coordAtoms[n][0],coordAtoms[n][1],
                       label=typeAtoms[n], c="red")
        n += 1

    #Add the legend
    ax.legend()
    plt.savefig('C:/File_name.jpeg') #User file path
    #Plotting the data

S7=[] #The list of atomic pair types
R=[] #The list of interatomic distances
K=[] #The list of bonded atoms
for h in range(0,len(S4)):
    for v in range(0,len(S4[h])):
        if round(((S4[h][v]-10)**2)*(13.28276-
(17.08506*S4[h][v])+(8.262515*(S4[h][v]**2))),3) in U:
            S7.append('Ni-Ni')
            R.append(S4[h][v])
            K.append(S2[h])
            K.append(S2[v])
        else:
            None
        if round(((S4[h][v]-10)**2)*(29.1429813-
(23.397503*S4[h][v])+(4.7578297*(S4[h][v]**2))),3) in U:
            S7.append('Cr-Cr')
            R.append(S4[h][v])
            K.append(S2[h])
            K.append(S2[v])
        else:
            None
        if round(((S4[h][v]-10)**2)*(26.27034-
(24.40109*S4[h][v])+(6.957871*(S4[h][v]**2))),3) in U:
            S7.append('Fe-Fe')
            R.append(S4[h][v])
            K.append(S2[h])
            K.append(S2[v])
        else:
            None
        if round(((0.850704*(S4[h][v]-10)**2)*(26.27034-
(24.40109*S4[h][v])+(6.957871*(S4[h][v]**2))))+((0.186420*(S4[h][v]-
```

```python
10)**2)*(29.1429813-(29.3975027*S4[h][v])+(4.7578297*(S4[h][v]**2)))),3) in U:
            S7.append('Fe-Cr')
            R.append(S4[h][v])
            K.append(S2[h])
            K.append(S2[v])
        else:
            None
        if round(((0.516721*(S4[h][v]-10)**2)*(26.27034-(24.40109*S4[h][v])+(6.957871*(S4[h][v]**2))))+((0.275781*(S4[h][v]-10)**2)*(13.28276-(17.08506*S4[h][v])+(8.262515*(S4[h][v]**2)))),3) in U:
            S7.append('Fe-Ni')
            R.append(S4[h][v])
            K.append(S2[h])
            K.append(S2[v])
        else:
            None
        if round(((0.80590965*(S4[h][v]-10)**2)*(13.28276-(17.08506*S4[h][v])+(8.262515*(S4[h][v]**2))))+((0.192553163985*(S4[h][v]-10)**2)*(29.1429813-(23.3975027*S4[h][v])+(4.7578297*(S4[h][v]**2)))),3) in U:
            S7.append('Ni-Cr')
            R.append(S4[h][v])
            K.append(S2[h])
            K.append(S2[v])
        else:
            None
print('Atomic (useful) pairs')
T7=datetime.datetime.now()
print(T7,'\n')
print('Atomic (useful) distances')
T8=datetime.datetime.now()
print(T8,'\n')
print('Atomic (useful) coordinates')
T9=datetime.datetime.now()
print(T9,'\n')

#======================= User output block and aftereffect
import os
path=r'C:\'
file='Radial distribution function.txt'
with open(os.path.join(path, file), 'w') as f:
    if 'Energy' not in file:
        f.write('Energy, eV\t\t\t'+'Distance, A\n\n')
    for i in range(0,len(U)):
        f.write(str(U[i])+'\t'+'\t'+str(R[i])+'\n')
f.close()
file='Atomic coordinates.txt'
with open(os.path.join(path, file), 'w') as f:
    if 'Coordinates' not in file:
        f.write('Coordinates\n\n')
    for i in range(0,len(K)):
        try:
            f.write(str(K[i])+'\n')
        except IndexError:
            None
print('The programme has done. Please, check the file directory', path)
T10=datetime.datetime.now()
print(T10)

MyPlot(S7,K)

os.system('shutdown /s /t 10') # Shutdown after N seconds
```

ii) *Parallel counting of model AMAG-225 assembling on arbitrary used CPU*

```python
from mpi4py import MPI
from sys import argv
import math
import random
import numpy as np
import copy
import time

#Launch the parallel programme counting

def f1(S2, start, end):
    S2 = list(S2)
    L = []
    for i in range(start, end):
        for n in range(0, len(S2)):
            L.append(round(math.sqrt(((S2[n][0]-S2[i][0])**2)+((S2[n][1]-S2[i][1])**2)), 3))
    return L

comm = MPI.COMM_WORLD # Binding to a processor group
rank = comm.Get_rank() # Number of a processor
nprocs = comm.Get_size() # Amount of processors

a1 = 0
b1 = 30
N1 = int(argv[1])
N2 = int(argv[2])
N3 = int(argv[3])
N = N1+N2+N3
S1 = []
S2 = []

#Launch the timer for the first count stage
start_time_step1 = MPI.Wtime()

#Creating the particles on zero processor
if rank == 0:
    for i in range(0, N):
        a = round(random.uniform(a1, b1), 3)
        b = round(random.uniform(a1, b1), 3)
        S1.append(a)
        S1.append(b)
    for i in range(0, len(S1), 2):
        S2.append(S1[i:i+2])

#Creating the lists with the further processor shifting
if rank == 0:
    #Count: the size of each sub-task
    ave, res = divmod(len(S2), nprocs)
    count = [(ave + 1) if p < res else ave for p in range(nprocs)]
    count = np.array(count)
    #Displacement: the starting index of each sub-task
    displ = [sum(count[:p] * len(S2)) for p in range(nprocs)]
    displ = np.array(displ)
```

```python
    else:
        #Initialize count on working processes
        count = np.zeros(nprocs, dtype = int)
        displ = np.zeros(nprocs, dtype = int)

    #Distribution between processors by amount
    comm.Bcast(count, root = 0)
    #Distribution of the lists between processors
    S2 = comm.bcast(S2, root = 0)
    #Calling of the f1 function on every processor
    displ_rank = [sum(count[:p]) for p in range(nprocs)]
    L = f1(S2, displ_rank[rank], displ_rank[rank] + count[rank])

    #Get the result on zero processor from others
    sendbuf1 = np.array(L)
    bufSTEP1 = np.zeros(len(S2) * len(S2))
    count = [x * len(S2) for x in list(count)]
    count = np.array(count)
    comm.Gatherv(sendbuf1, [bufSTEP1, count, displ, MPI.LONG_DOUBLE], root = 0)

    #Stop timer for the first count stage
    end_time_step1 = MPI.Wtime()
    #Measured time
    time_step1 = end_time_step1 - start_time_step1
    #Measured times from all processors are sent to zero unit
    total_time_step1 = comm.reduce(time_step1, op = MPI.SUM, root = 0)
    if rank == 0:
        print(f"Total time for step 1: {total_time_step1} seconds")

    #Synchronisation
    MPI.COMM_WORLD.Barrier()
    S4 = []
    S3 = []
    S5 = []
    B1 = []

    #Launch the timer for the second count stage
    start_time_step2 = MPI.Wtime()

    #Calculations on zero processor
    if rank == 0:
        for i in range(0, bufSTEP1.size, len(S2)):
            S4.append(list(bufSTEP1[i:i+len(S2)]))
        for i in range(0, len(S4)):
            for k in range(0, len(S4[i])):
                if S4[i][k] in S5:
                    S4[i][k] = 0
                else:
                    S5.append(S4[i][k])
        for i in range(0, len(S4)):
            for k in range(0, len(S4[i])):
                if S4[i][k] < 2.7:
                    S4[i][k] = 0
                else:
                    None
        B1 = copy.deepcopy(S4)
        for i in range(0, len(B1)):
            for j in range(0, len(B1[i])):
                B1[i][j] = 0
    def f2(S4, B1, start, end):
        M1 = []
        C1 = N1
```

```python
        C2 = N2
        C3 = N3
        for i in range(start, end):
            for k in range(0, len(S4[i])):
                if 2.7 <= S4[i][k] <= 10:
                    if 2.8 <= S4[i][k] <= 10 and C3 >= 2:
                        M1.append(((S4[i][k]-10)**2)*(29.1429813-(23.397503*S4[i][k])+(4.7578297*(S4[i][k]**2))))
                    else:
                        M1.append(0)
                    if 2.8 <= S4[i][k] <= 10 and C1 >= 2:
                        M1.append(((S4[i][k]-10)**2)*(26.27034-(24.40109*S4[i][k])+(6.957871*(S4[i][k]**2))))
                    else:
                        M1.append(0)
                    if 2.7 <= S4[i][k] <= 10 and C2 >= 2:
                        M1.append(((S4[i][k]-10)**2)*(13.28276-(17.08506*S4[i][k])+(8.262515*(S4[i][k]**2))))
                    else:
                        M1.append(0)
                    if 2.8 <= S4[i][k] <= 10 and C1 >= 1 and C3 >= 1:
                        M1.append(((0.850704*(S4[i][k]-10)**2)*(26.27034-(24.40109*S4[i][k])+(6.957871*(S4[i][k]**2))))+((0.186420*(S4[i][k]-10)**2)*(29.1429813-(29.3975027*S4[i][k])+(4.7578297*(S4[i][k]**2)))))
                    else:
                        M1.append(0)
                    if 2.75 <= S4[i][k] <= 10 and C1 >= 1 and C2 >= 1:
                        M1.append(((0.516721*(S4[i][k]-10)**2)*(26.27034-(24.40109*S4[i][k])+(6.957871*(S4[i][k]**2))))+((0.275781*(S4[i][k]-10)**2)*(13.28276-(17.08506*S4[i][k])+(8.262515*(S4[i][k]**2)))))
                    else:
                        M1.append(0)
                    if 2.75 <= S4[i][k] <= 10 and C2 >= 1 and C3 >= 1:
                        M1.append(((0.80590965*(S4[i][k]-10)**2)*(13.28276-(17.08506*S4[i][k])+(8.262515*(S4[i][k]**2))))+((0.192553163985*(S4[i][k]-10)**2)*(29.1429813-(23.3975027*S4[i][k])+(4.7578297*(S4[i][k]**2)))))
                    else:
                        M1.append(0)
                    M2 = []
                    for j in range(0, len(M1)):
                        if M1[j] != 0:
                            M2.append(M1[j])
                        else:
                            None
                    M1 = M2
                    try:
                        if len(set(M1)) == len(M1):
                            None
                        else:
                            print('Duplicates are detected!')
                            exit()
                    except ValueError:
                        None
                    try:
                        B1[i][k] = min(M1)
                        if min(M1) == ((S4[i][k]-10)**2)*(29.1429813-(23.397503*S4[i][k])+(4.7578297*(S4[i][k]**2))):
                            C3 = C3-2
                        elif min(M1) == ((S4[i][k]-10)**2)*(26.27034-(24.40109*S4[i][k])+(6.957871*(S4[i][k]**2))):
                            C1 = C1-2
                        elif min(M1) == ((S4[i][k]-10)**2)*(13.28276-(17.08506*S4[i][k])+(8.262515*(S4[i][k]**2))):
```

```python
                                    C2 = C2-2
                                elif min(M1) == ((0.850704*(S4[i][k]-10)**2)*(26.27034-(24.40109*S4[i][k])+(6.957871*(S4[i][k]**2))))+((0.186420*(S4[i][k]-10)**2)*(29.1429813-(29.3975027*S4[i][k])+(4.7578297*(S4[i][k]**2)))):
                                    C1 = C1-1
                                    C3 = C3-1
                                elif min(M1) == ((0.516721*(S4[i][k]-10)**2)*(26.27034-(24.40109*S4[i][k])+(6.957871*(S4[i][k]**2))))+((0.275781*(S4[i][k]-10)**2)*(13.28276-(17.08506*S4[i][k])+(8.262515*(S4[i][k]**2)))):
                                    C1 = C1-1
                                    C2 = C2-1
                                elif min(M1) == ((0.80590965*(S4[i][k]-10)**2)*(13.28276-(17.08506*S4[i][k])+(8.262515*(S4[i][k]**2))))+((0.192553163985*(S4[i][k]-10)**2)*(29.1429813-(23.3975027*S4[i][k])+(4.7578297*(S4[i][k]**2)))):
                                    C2 = C2-1
                                    C3 = C3-1
                                else:
                                    None
                        except ValueError:
                            None
                    M1 = []
                    if B1[i][k] != 0:
                        break
                else:
                    B1[i][k] = 0
    return B1[start: end]
#Creating the lists with the further processor shifting
if rank == 0:
    #Count: the size of each sub-task
    ave, res = divmod(len(S4), nprocs)
    count = [(ave + 1) if p < res else ave for p in range(nprocs)]
    count = np.array(count)
    #Displacement: the starting index of each sub-task
    displ = [sum(count[:p] * len(S4)) for p in range(nprocs)]
    displ = np.array(displ)
else:
    #Initialize count on working processes
    count = np.zeros(nprocs, dtype = int)
    displ = np.zeros(nprocs, dtype = int)

#Distribution between processors by amount
comm.Bcast(count, root = 0)
#Distribution of the S4 list from zero processor between others
S4 = comm.bcast(S4, root = 0)
#Distribution of the B1 list from zero processor between others
B1 = comm.bcast(B1, root = 0)

#Calling of the f2 function on every processor
displ_rank = [sum(count[:p]) for p in range(nprocs)]
L = f2(S4, B1, displ_rank[rank], displ_rank[rank] + count[rank])

#Get the result on zero processor from others
sendbuf2 = np.array(L)
bufSTEP2 = np.zeros(len(S4) * len(S4))
count = [x * len(S4) for x in list(count)]
count = np.array(count)
comm.Gatherv(sendbuf2, [bufSTEP2, count, displ, MPI.LONG_DOUBLE], root = 0)

bufSTEP2_modify = []

#Output data from the second stage on zero processor
if rank == 0:
    #Slicing of the list by S4 length
```

```python
        buf = np.array_split(bufSTEP2, len(S4))
        for x in buf:
            bufSTEP2_modify.append(list(x))

end_time_step2 = MPI.Wtime()
#Stop timer for the second count stage
time_step2 = end_time_step2 - start_time_step2
#Measured times from all processors are sent to zero unit
total_time_step2 = comm.reduce(time_step2, op = MPI.SUM, root = 0)
if rank == 0:
    print(f"Total time for step 2: {total_time_step2} seconds")

def f3(P, start, end):
    U = []
    for i in range(start, end):
        B2 = []
        for k in range(0, len(P[i])):
            if P[i][k] != 0 and P[i][k] not in U:
                B2.append(P[i][k])
            else:
                None
        try:
            U.append(min(B2))
        except ValueError:
            None
        for l in range(start, end):
            for m in range(0, len(P[l])):
                if P[l][m] in U:
                    k = P[l].index(P[l][m])
                    P[k] = [0]
                else:
                    None
    return U

#Launch the timer for the third count stage
start_time_step3 = MPI.Wtime()
#Creating the lists with the further processor shifting
if rank == 0:
    #Count: the size of each sub-task
    ave, res = divmod(len(bufSTEP2_modify), nprocs)
    count = [(ave + 1) if p < res else ave for p in range(nprocs)]
    count = np.array(count)
else:
    #Initialize count on working processes
    count = np.zeros(nprocs, dtype = int)
#Distribution between processors by amount
comm.Bcast(count, root = 0)
#Distribution of the bufSTEP2_modify list from zero processor between others
P = comm.bcast(bufSTEP2_modify, root = 0)
#Calling of the f3 function on every processor
displ_rank = [sum(count[:p]) for p in range(nprocs)]
U = f3(P, displ_rank[rank], displ_rank[rank] + count[rank])
Full_U = []
if rank == 0:
    Full_U = U
    for p in range(1, nprocs):
        U = comm.recv(source = p)
        Full_U = Full_U + U
else:
    comm.send(U, dest = 0)

end_time_step3 = MPI.Wtime()
#Stop timer for the third count stage
```

```python
    time_step3 = end_time_step3 - start_time_step3
    #Measured times from all processors are sent to zero unit
    total_time_step3 = comm.reduce(time_step3, op = MPI.SUM, root = 0)
    if rank == 0:
        print(f"Total time for step 3: {total_time_step3} seconds")

def f4(S4, U, start, end):
    S7 = []
    R = []
    K = []
    for h in range(start, end): # 0, len(S4)
        for v in range(0, len(S4[h])):
            if ((S4[h][v]-10)**2)*(13.28276-(17.08506*S4[h][v])+(8.262515*(S4[h][v]**2))) in U:
                S7.append('Ni-Ni')
                R.append(S4[h][v])
                K.append(S2[h])
                K.append(S2[v])
                continue
            elif ((S4[h][v]-10)**2)*(29.1429813-(23.397503*S4[h][v])+(4.7578297*(S4[h][v]**2))) in U:
                S7.append('Cr-Cr')
                R.append(S4[h][v])
                K.append(S2[h])
                K.append(S2[v])
                continue
            elif ((S4[h][v]-10)**2)*(26.27034-(24.40109*S4[h][v])+(6.957871*(S4[h][v]**2))) in U:
                S7.append('Fe-Fe')
                R.append(S4[h][v])
                K.append(S2[h])
                K.append(S2[v])
                continue
            elif ((0.850704*(S4[h][v]-10)**2)*(26.27034-(24.40109*S4[h][v])+(6.957871*(S4[h][v]**2))))+((0.186420*(S4[h][v]-10)**2)*(29.1429813-(29.3975027*S4[h][v])+(4.7578297*(S4[h][v]**2)))) in U:
                S7.append('Fe-Cr')
                R.append(S4[h][v])
                K.append(S2[h])
                K.append(S2[v])
                continue
            elif ((0.516721*(S4[h][v]-10)**2)*(26.27034-(24.40109*S4[h][v])+(6.957871*(S4[h][v]**2))))+((0.275781*(S4[h][v]-10)**2)*(13.28276-(17.08506*S4[h][v])+(8.262515*(S4[h][v]**2)))) in U:
                S7.append('Fe-Ni')
                R.append(S4[h][v])
                K.append(S2[h])
                K.append(S2[v])
                continue
            elif ((0.80590965*(S4[h][v]-10)**2)*(13.28276-(17.08506*S4[h][v])+(8.262515*(S4[h][v]**2))))+((0.192553163985*(S4[h][v]-10)**2)*(29.1429813-(23.3975027*S4[h][v])+(4.7578297*(S4[h][v]**2)))) in U:
                S7.append('Ni-Cr')
                R.append(S4[h][v])
                K.append(S2[h])
                K.append(S2[v])
                continue
            else:
                None

    return [S7,R,K]
#Launch the timer for the fourth count stage
start_time_step4 = MPI.Wtime()
```

```python
#Creating the lists with the further processor shifting
if rank == 0:
    #Count: the size of each sub-task
    ave, res = divmod(len(S4), nprocs)
    count = [(ave + 1) if p < res else ave for p in range(nprocs)]
    count = np.array(count)
else:
    #Initialize count on working processes
    count = np.zeros(nprocs, dtype = int)
#Distribution between processors by amount
comm.Bcast(count, root = 0)
# Distribution of the S4 list from zero processor between others
S4 = comm.bcast(S4, root = 0)
# Distribution of the Full_U list from zero processor between others
Full_U = comm.bcast(Full_U, root = 0)
#Calling of the f4 function on every processor
displ_rank = [sum(count[:p]) for p in range(nprocs)]
Result = f4(S4, Full_U, displ_rank[rank], displ_rank[rank] + count[rank])
Full_Result = []
if rank == 0:
    Full_Result = Result
    for p in range(1, nprocs):
        Result = comm.recv(source = p)
        if len(Full_Result) == len(Result):
            for i in range(0, len(Result)):
                if len(Result[i]) != 0:
                    Full_Result[i] = Full_Result[i] + Result[i]
else:
    comm.send(Result, dest = 0)
# Stop timer for the fourth count stage
end_time_step4 = MPI.Wtime()
time_step4 = end_time_step4 - start_time_step4
#Measured times from all processors are sent to zero unit
total_time_step4 = comm.reduce(time_step4, op = MPI.SUM, root = 0)
if rank == 0:
    print(f"Total time for step 4: {total_time_step4} seconds")

# -----(OUTPUT DATA)------ #
# Output data on zero processor
if rank == 0:
   #Launch the timer for the fifth count stage
   start_time_step5 = MPI.Wtime()
   print('Energy, eV\t\t\t' + 'Coordinates[x,y]\t\t\t' + 'Distance, A\n')
   for i in range(0, len(Full_U)):
       print(str(Full_U[i]) + '\t' + '\t' + '\t' + str(Full_Result[2][i]) +
'\t' + '\t' + '\t' + str(Full_Result[1][i]) + '\n')
   print('Calculation success!')
   print('\n' + '\n' + 'Debugging:' + '\n' + 'array = ' +
str(bufSTEP2_modify))
# Stop timer for the fifth count stage
   end_time_step5 = MPI.Wtime()
   time_step5 = end_time_step5 - start_time_step5
   print(f"Total time for step 5: {time_step5} seconds")
```